\documentclass{aip-cp}

\usepackage{graphicx}
\usepackage{grffile}
\usepackage{savesym}  
\savesymbol{iint}
\savesymbol{iiint}
\savesymbol{iiiint}
\savesymbol{idotsint}
\usepackage[%
  sumlimits,
  intlimits,
  namelimits]{amsmath}
\restoresymbol{AMS}{iint}  
\restoresymbol{AMS}{iiint}
\restoresymbol{AMS}{iiiint}
\restoresymbol{AMS}{idotsint}
\usepackage{mathtools}
\usepackage{amssymb}
\usepackage{amsfonts}
\usepackage{amsxtra}
\usepackage[numbers]{natbib}
\usepackage{textcomp}

\makeatletter
\g@addto@macro\bfseries{\boldmath}
\makeatother

\newcommand*{\etaprime}{\eta^{(}{'}\vphantom{\eta}^{)}}

\begin{document}

\title{Recent Results on Spectroscopy from COMPASS}

\author[aff1]{Boris Grube\corref{cor1}}
\author{for the COMPASS Collaboration}
\affil[aff1]{Physik-Department E18, Technische Universit\"at M\"unchen, Garching, Germany}
\corresp[cor1]{Corresponding author: bgrube@tum.de}

\maketitle

\begin{abstract}
  The COmmon Muon and Proton Apparatus for Structure and Spectroscopy
  (COMPASS) is a multi-purpose fixed-target experiment at the CERN
  Super Proton Synchrotron (SPS) aimed at studying the structure and
  spectrum of hadrons.  The two-stage spectrometer has a good
  acceptance for charged as well as neutral particles over a wide
  kinematic range and is thus able to measure a wide range of
  reactions.  Light mesons are studied with negative (mostly $\pi^-$)
  and positive ($p$, $\pi^+$) hadron beams with a momentum of
  190~GeV/$c$.  The light-meson spectrum is investigated in various
  final states produced in diffractive dissociation reactions at
  squared four-momentum transfers to the target between 0.1~and
  1.0~$(\text{GeV}/c)^2$.  The flagship channel is the
  $\pi^-\pi^+\pi^-$ final state, for which COMPASS has recorded the
  currently largest data sample.  These data not only allow for
  measuring the properties of known resonances with high precision,
  but also for searching for new states.  Among these is a new
  resonance-like signal, the $a_1(1420)$, with unusual properties.
  The findings are confirmed by the analysis of the $\pi^-\pi^0\pi^0$
  final state.  Possible bias introduced by the parametrizations used
  to describe the $\pi\pi$ $S$-wave is studied using a novel analysis
  technique, which extracts the amplitude of the $\pi^+\pi^-$
  sub-system as a function of $3\pi$ mass from the data.  Of
  particular interest is the resonance content of the partial wave
  with spin-exotic $J^{PC} = 1^{-+}$ quantum numbers, which are
  forbidden for quark-antiquark states.  This wave is studied in the
  two $3\pi$ channels.  Further insight is gained by studying
  diffractively produced $\pi^-\eta$ or $\pi^-\eta'$ final states.
\end{abstract}

\section{INTRODUCTION}

The COMPASS experiment~\cite{compass} has recorded large data sets of
diffractive dissociation reactions of the type
$\pi^- + p \to X^- + p_\text{recoil}$ using a 190~GeV/$c$ pion beam on
a liquid-hydrogen target.  In this process, the beam hadron is excited
to some intermediate state $X^-$ via $t$-channel Reggeon exchange with
the target.  At 190~GeV/$c$ beam momentum pomeron exchange is
dominant.  Diffractive reactions are known to exhibit a rich spectrum
of intermediate states $X^-$.  In the past, several candidates for
spin-exotic mesons have been reported in pion-induced
diffraction~\cite{exotic_1,exotic_2}.  The mesonic states $X^-$ decay
into various multi-hadron final states, which are detected by the
spectrometer.  Here, we consider the $\pi^-\eta$, $\pi^-\eta'$,
$\pi^-\pi^+\pi^-$, and $\pi^-\pi^0\pi^0$ channels.  In order to
disentangle the different contributing intermediate states $X^-$, for
each final state a partial-wave analysis (PWA) is performed.  In
addition to the final-state particles from the $X^-$ decays, the
recoiling proton is also measured.  This helps to suppress backgrounds
and ensures an exclusive measurement by applying energy and momentum
conservation in the event selection.

The scattering process is characterized by two kinematic variables:
the squared total center-of-mass energy $s$ and the squared
four-momentum transfer to the target
$t = (p_\text{beam} - p_{X})^2 < 0$.  It is customary to use the
reduced four-momentum transfer $t' \equiv |t| - |t|_\text{min}$
instead of $t$, where $|t|_\text{min}$ is the minimum value of $|t|$
for a given invariant mass of $X^-$.  The analyses presented are all
performed in the range $0.1 < t' < 1.0~(\text{GeV}/c)^2$.

\section{PARTIAL-WAVE ANALYSIS OF THE $\pi^-\eta$ AND $\pi^-\eta'$ FINAL STATES}

The $\pi^-\eta$ and $\pi^-\eta'$ final states are interesting, because
in both channels the odd-spin partial waves have spin-exotic $J^{PC}$
quantum numbers that are forbidden for quark-antiquark states in the
non-relativistic limit.  By convention, the $C$-parity is that of the
neutral partner of the $X^-$ in the isospin triplet.  A comparison of
the two channels also gives insight into the role of flavor symmetry.
The $\eta$ is reconstructed via its decay into $\pi^+\pi^-\pi^0$,
where $\pi^0 \to \gamma\gamma$.  For the $\eta'$, its decay into
$\pi^+\pi^-\eta$ with $\eta \to \gamma\gamma$ is used.

The partial-wave analysis of the two channels~\cite{compass_eta_pi} is
performed in 40~MeV/$c^2$ wide bins of the final-state invariant mass
and makes therefore no assumptions on the resonance content of the
partial waves.  The two-body decay of the $X^-$ is described in the
Gottfried-Jackson frame by spherical harmonics.  The partial waves are
defined by the orbital angular momentum $L$ between the pion and the
$\etaprime$, which corresponds to the spin $J$ of the $X^-$, by the
absolute value $M$ of the spin projection, and by the reflectivity
$\epsilon = \pm 1$, which corresponds to the naturality of the
exchanged particle.  Partial waves with $\epsilon = -1$ are found to be
negligibly small, except for the $S$ wave with $M = 0$, which
contributes 0.5\% in the $\pi^-\eta$ and 1.1\% in the $\pi^-\eta'$
channel.  This is in agreement with the expected pomeron dominance at
high energies.  For positive reflectivity (naturality), only waves
with $M \geq 1$ are allowed so that $L = J \geq 1$.  The data are
dominated by $M = 1$ partial waves.  The PWA model includes $M = 1$
waves with $L = 1$ to 6.  Higher $M$ states are suppressed.  Only the
$\pi^-\eta$ $D$ wave shows significant intensity for $M = 2$.

\begin{figure}[!tb]
  \centering
  \begin{minipage}{\textwidth}
    \centering
    \includegraphics[width=0.495\textwidth]{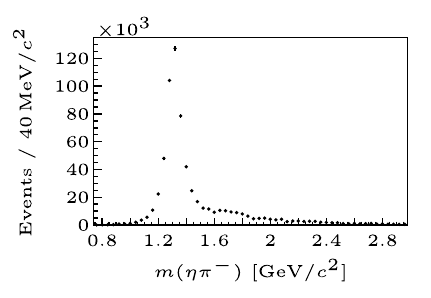}
    \includegraphics[width=0.495\textwidth]{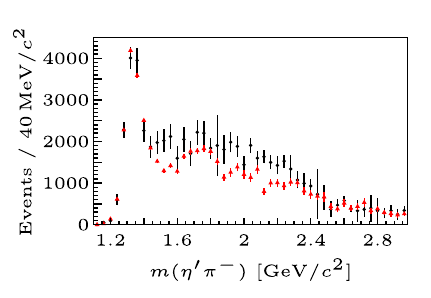} \\
    \includegraphics[width=0.495\textwidth]{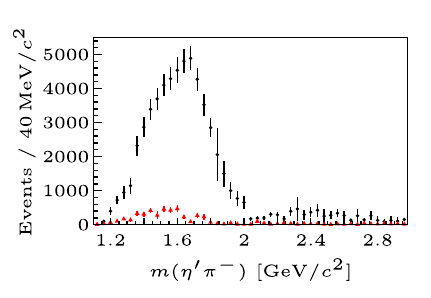}
    \includegraphics[width=0.495\textwidth]{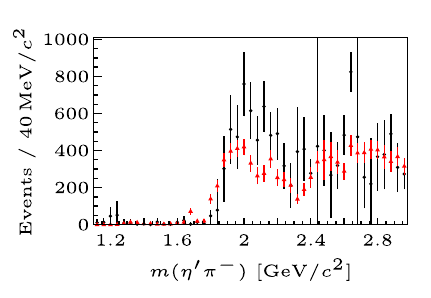}
  \end{minipage}
  \caption{Top left: intensity of the $D$ wave with $J^{PC} = 2^{++}$
    in the $\pi^-\eta$ final state.  The other three panels show in
    black the intensities of the $2^{++}$ ($D$ wave; top right),
    $1^{-+}$ ($P$ wave; bottom left), and $4^{++}$ ($G$ wave; bottom
    right) $\pi^-\eta'$ waves.  The red data points represent the
    corresponding $\pi^-\eta$ intensities scaled by
    Eq.~\eqref{eq:ps_scaling}. Figures from
    Ref.~\cite{compass_eta_pi}.}
  \label{fig:eta_pi}
\end{figure}

The $\pi^-\eta$ data are dominated by the $D$ wave with a prominent
peak corresponding to the $a_2(1320)$ (see Fig.~\ref{fig:eta_pi} top
left).  In contrast, the $D$ wave is much weaker in the $\pi^-\eta'$
channel (black data points in Fig.~\ref{fig:eta_pi} top right).
However, in the quark-line picture and assuming point-like resonances,
the $\pi^-\eta$ and $\pi^-\eta'$ intensities for a partial wave with
spin $J$ should be related by the ratios of phase space and
angular-momentum barrier factors and by the branching-fraction ratio
$B$ for $\eta$ and $\eta'$ decaying to $\pi^+\pi^-\gamma\gamma$:
\begin{equation}
  \label{eq:ps_scaling}
  N_J^{\pi\eta'}(m) \propto
  B\, \left[ \frac{q^{\pi\eta'}(m)}{q^{\pi\eta}(m)} \right]^{2J + 1}\!\! N_J^{\pi\eta}(m)
\end{equation}
Here $N_J$ is the acceptance-corrected number of events in the partial
wave with spin $J$ of the respective final state and $q$ is the
corresponding breakup momentum.

Scaling the $\pi^-\eta$ partial-wave intensities according to
Eq.~\eqref{eq:ps_scaling} (red data points in Fig.~\ref{fig:eta_pi})
yields for even-spin waves intensity distributions that are quite
similar to those in the $\pi^-\eta'$ channel (see
Fig.~\ref{fig:eta_pi} right column).  This is consistent with the
assumption that in both channels the intermediate states $X^-$ couple
to the same final-state flavor content.  It is interesting to note
that the similarities in the even-spin waves are not limited to the
resonance regions, i.e. $a_2(1320)$ in the $D$ and $a_4(2040)$ in the
$G$ wave.  Similar intensity distributions are also observed in the
high-mass regions, which contain large non-resonant contributions.
The picture is completely different for the spin-exotic odd-spin
partial waves.  The scaled $\pi^-\eta$ partial-wave intensities are
suppressed by factors of 5 to 10 compared to $\pi^-\eta'$.  This is
illustrated in Fig.~\ref{fig:eta_pi} bottom left for the
$J^{PC} = 1^{-+}$ wave, which is the dominant wave in the $\pi^-\eta'$
channel.

Resonance-model fits using relativistic Breit-Wigner amplitudes yield
resonance parameters for the $a_2(1320)$ and the $a_4(2040)$ that are
consistent with PDG values~\cite{pdg}.  From these fits the following
branching-fraction ratios are extracted:
\begin{equation*}
  B_2 \equiv \frac{N(a_2 \to \pi^-\eta')}{N(a_2 \to \pi^-\eta)} = (5 \pm 2)\%
  \qquad\text{and}\qquad
  B_4 \equiv \frac{N(a_4 \to \pi^-\eta')}{N(a_4 \to \pi^-\eta)} = (23 \pm 7)\%
\end{equation*}
Here the $N$ are the integrated Breit-Wigner intensities for the
respective decays.  The ratio $B_2$ is in agreement with previous
measurements; $B_4$ is measured for the first time.  For the
spin-exotic $P$ waves, the results of the resonance-model fit are
strongly model dependent.  Applying the models of previous
analyses~\cite{e852_eta_pi_1,e852_eta_pi_2,ves_eta_pi} yields
similar resonance parameters for the $\pi_1(1400)$ in $\pi^-\eta$ and
for the $\pi_1(1600)$ in $\pi^-\eta'$.  However, the results depend on
the modeling of the $D$ waves above the $a_2(1320)$ and on the
parametrization of the non-resonant component in the $P$ waves.  Both
have to be better understood before firm conclusions can be drawn
about the resonance content of the $P$ waves.

\section{PARTIAL-WAVE ANALYSIS OF THE $\pi^-\pi^+\pi^-$ AND $\pi^-\pi^0\pi^0$ FINAL STATES}

\begin{figure}[!tb]
  \centering
  \includegraphics[width=0.45\textwidth]{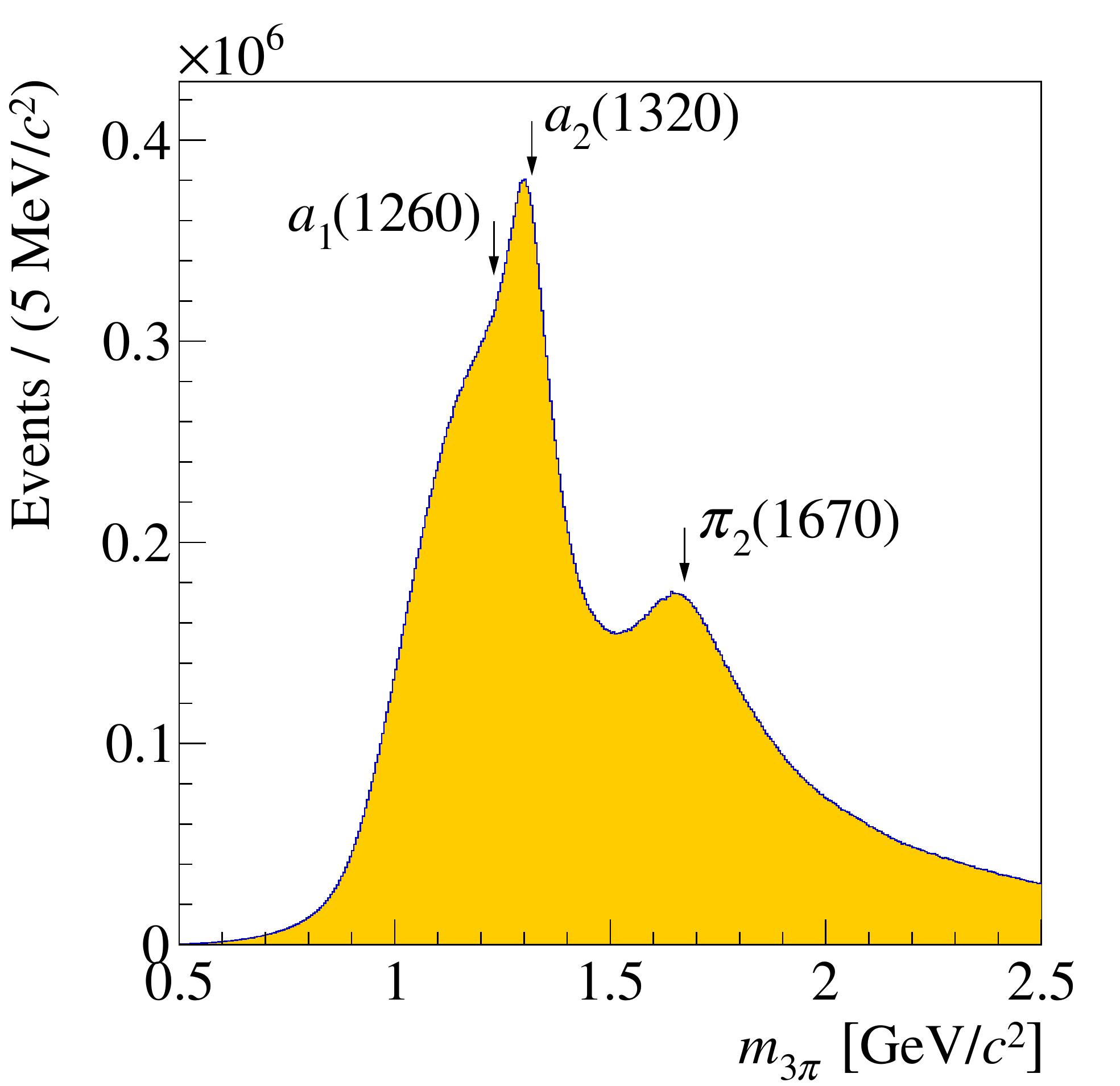}
  \includegraphics[width=0.45\textwidth]{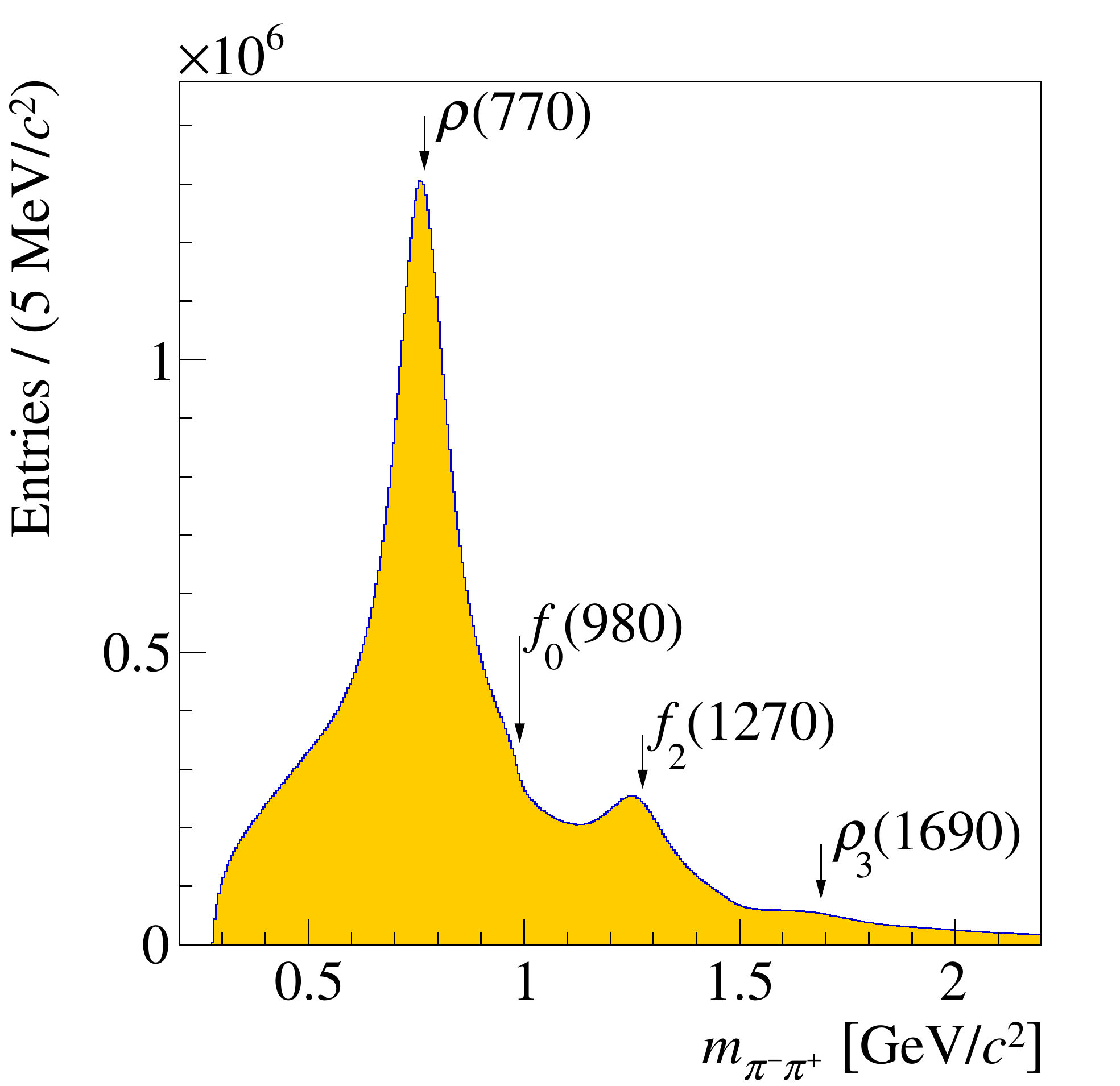}
  \caption{Left: $\pi^-\pi^+\pi^-$ invariant mass spectrum in the
    analyzed range; Right: invariant mass distribution of the
    $\pi^+\pi^-$ subsystem (two entries per event). Figures from
    Ref.~\cite{long_paper}.}
  \label{fig:mass}
\end{figure}

After all the selection cuts, the $3\pi$ data samples consist of
$46 \times 10^6$ $\pi^-\pi^+\pi^-$ and $3.5 \times 10^6$
$\pi^-\pi^0\pi^0$ exclusive events in the analyzed kinematic region of
three-pion mass, $0.5 < m_{3\pi} < 2.5~\text{GeV}/c^2$.
Figure~\ref{fig:mass} shows the $\pi^-\pi^+\pi^-$ mass spectrum
together with that of the $\pi^+\pi^-$ subsystem.  The known pattern
of resonances $a_1(1260)$, $a_2(1320)$, and $\pi_2(1670)$ is seen in
the $3\pi$ system along with $\rho(770)$, $f_0(980)$, $f_2(1270)$, and
$\rho_3(1690)$ in the $\pi^+\pi^-$ subsystem.

\subsection{Partial-Wave Decomposition}

The PWA of the $(3\pi)^-$ final states is based on the isobar model,
which assumes that the $X^-$ decays first into an intermediate
resonance, which is called the isobar, and a ``bachelor'' pion
($\pi^-$ for the $\pi^-\pi^+\pi^-$ final state; $\pi^0$ for
$\pi^-\pi^0\pi^0$).  In a second step, the isobar decays into two
pions.  In accordance with the $\pi^+\pi^-$ invariant mass spectrum
shown in Fig.~\ref{fig:mass} right and with analyses by previous
experiments, we include $[\pi\pi]_S$, $\rho(770)$, $f_0(980)$,
$f_2(1270)$, $f_0(1500)$, and $\rho_3(1690)$ as isobars into the fit
model.  Here, $[\pi\pi]_S$ represents the broad component of the
$\pi\pi$ $S$-wave.  Based on the six isobars, we have constructed a
set of partial waves that consists of 88~waves in total, including one
non-interfering isotropic wave representing three uncorrelated pions.
This constitues the largest wave set ever used in an analysis of the
$3\pi$ final state.  The partial-wave decomposition is performed in
narrow bins of the $3\pi$ invariant mass.  Since the data show a
complicated correlation of the $m_{3\pi}$ and $t'$ spectra, each
$m_{3\pi}$ bin is further subdivided into non-equidistant bins in the
four-momentum transfer $t'$.  For the $\pi^-\pi^+\pi^-$ channel
11~bins are used, for the $\pi^-\pi^0\pi^0$ final state 8~bins.  With
this additional binning in $t'$, the dependence of the partial-wave
amplitudes on the four-momentum transfer can be studied in detail.
The details of the analysis model are described in
Ref.~\cite{long_paper}.

The partial-wave amplitudes are extracted from the data as a function
of $m_{3\pi}$ and $t'$ by fitting the five-dimensional kinematic
distributions of the outgoing three pions.  The amplitudes do contain
information not only about the partial-wave intensities, but also
about the relative phases of the partial waves.  The latter are
crucial for resonance extraction.  Three-pion partial waves are
defined by the quantum numbers of the $X^-$ (spin $J$, parity $P$,
$C$-parity, absolute value $M$ of the spin projection), the naturality
$\epsilon = \pm 1$ of the exchange particle, the isobar, and the
orbital angular momentum $L$ between the isobar and the bachelor pion.
These quantities are summarized in the partial-wave notation
$J^{PC}\,M^\epsilon\,\text{[isobar]}\,\pi\,L$.  Since at the used beam
energies pomeron exchange is dominant, 80 of the 88~partial waves in
the model have $\epsilon = +1$.

\subsection{The $J^{PC} = 1^{-+}$ Spin-Exotic Wave}

\begin{figure}[!tb]
  \centering
  \includegraphics[width=0.45\textwidth]{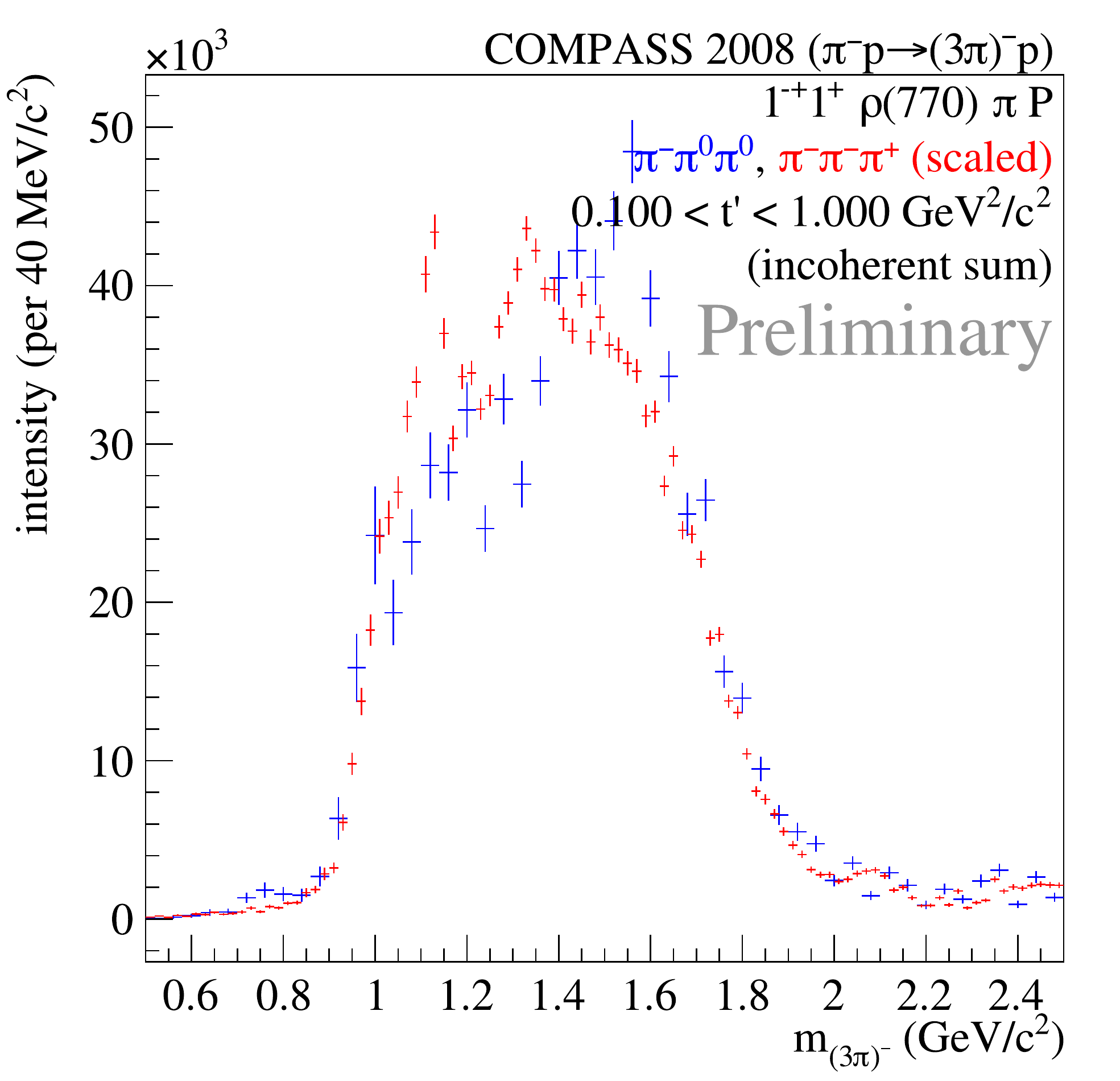}
  \raisebox{-0.9ex}{\includegraphics[width=0.445\textwidth]{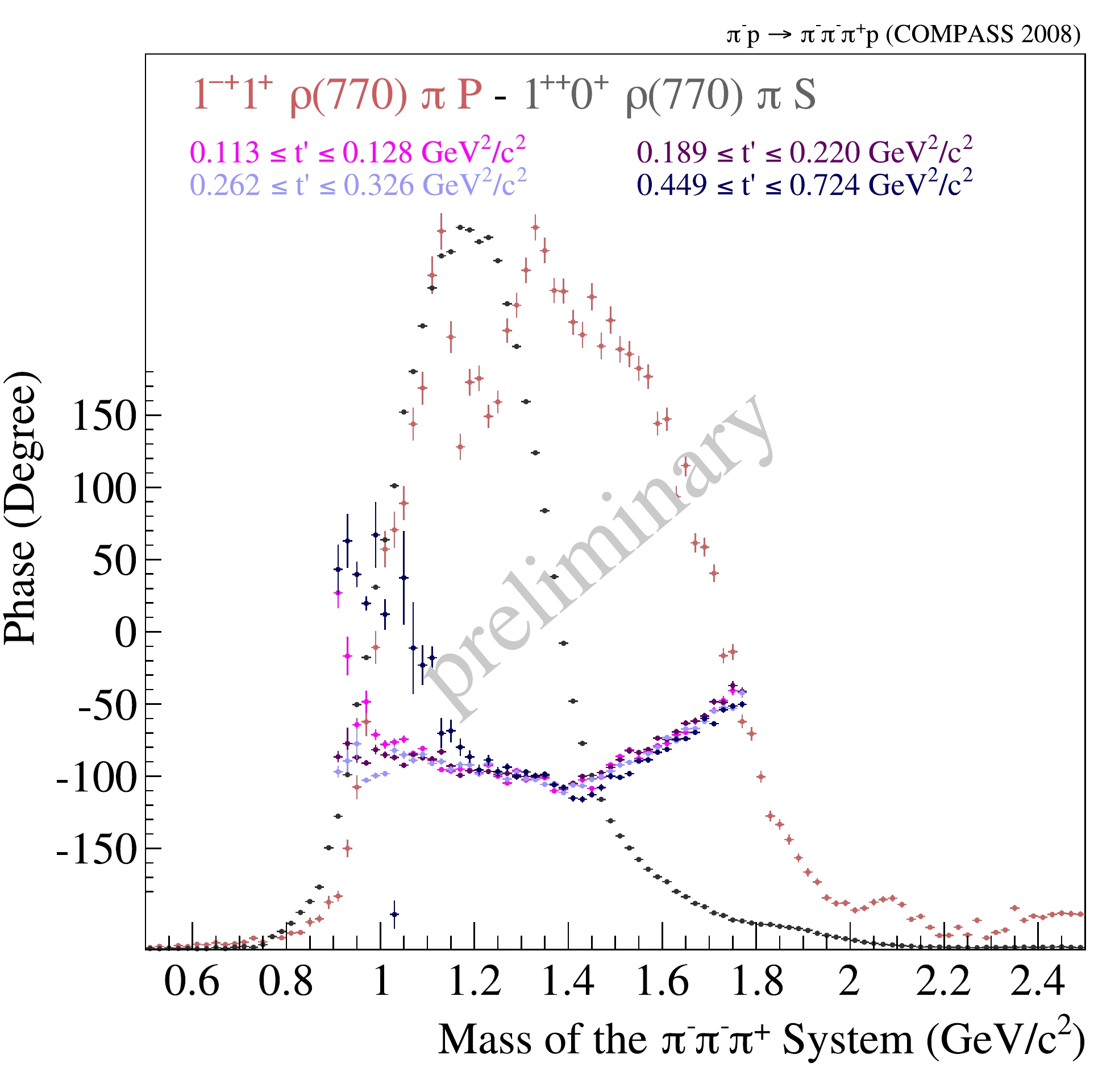}}
  \caption{Left: Intensity of the $1^{-+}\,1^+\,\rho(770)\,\pi\,P$
    wave summed over all $t'$ bins for the $\pi^-\pi^+\pi^-$ (red) and
    the $\pi^-\pi^0\pi^0$ (blue) final state.  Right: Phase of the
    $1^{-+}\,1^+\,\rho(770)\,\pi\,P$ wave relative to the
    $1^{++}\,0^+\,\rho(770)\,\pi\,S$ wave for four $t'$ bins in the
    $\pi^-\pi^+\pi^-$ data.  The black and dark red points show the
    shapes of the intensity distributions of the two waves scaled to
    the same height.}
  \label{fig:1mp}
\end{figure}

The 88-wave model also contains waves with spin-exotic $J^{PC}$
quantum numbers.  The most interesting of these waves is the
$1^{-+}\,1^+\,\rho(770)\,\pi\,P$ wave, which contributes less than 1\%
to the total intensity.  Previous analyses claimed a resonance, the
$\pi_1(1600)$, at about 1.6~GeV/$c^2$ in this
channel~\cite{bnl_1,compass_pb}.  Figure~\ref{fig:1mp} left shows the
intensity of this partial wave for the two final states
($\pi^-\pi^+\pi^-$ in red, $\pi^-\pi^0\pi^0$ in blue).  The two
distributions are scaled to have the same integral.  Both decay
channels are in fair agreement and exhibit a broad enhancement
extending from about 1.0 to 1.8~GeV/$c^2$ in $m_{3\pi}$.  In the 1.0
to 1.2~GeV/$c^2$ mass range the intensity depends strongly on the
details of the fit model.  Peak-like structures in this region are
probably due to imperfections of the applied PWA model.

\begin{figure}[!tb]
  \centering
  \includegraphics[width=0.45\textwidth]{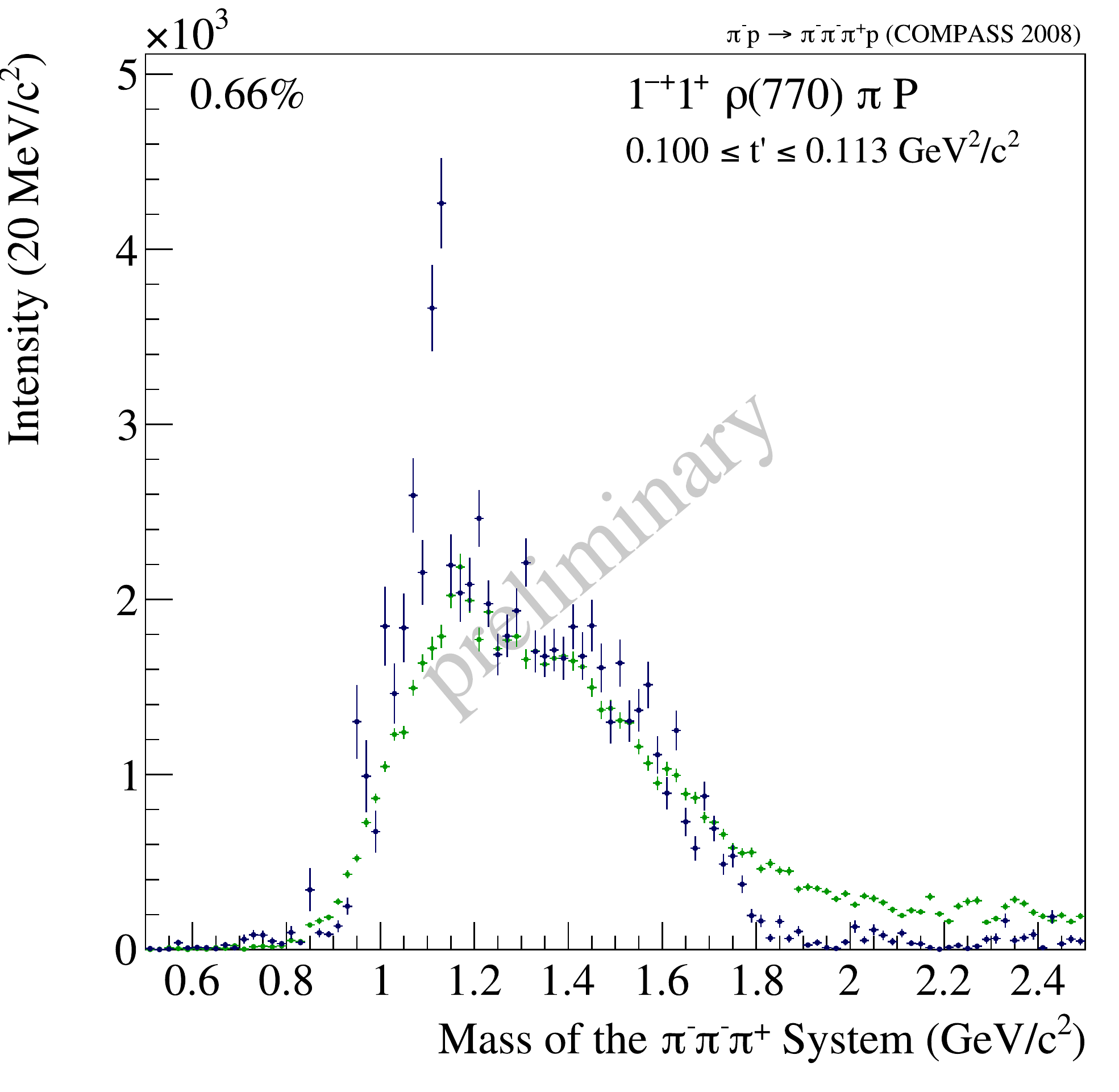}
  \includegraphics[width=0.45\textwidth]{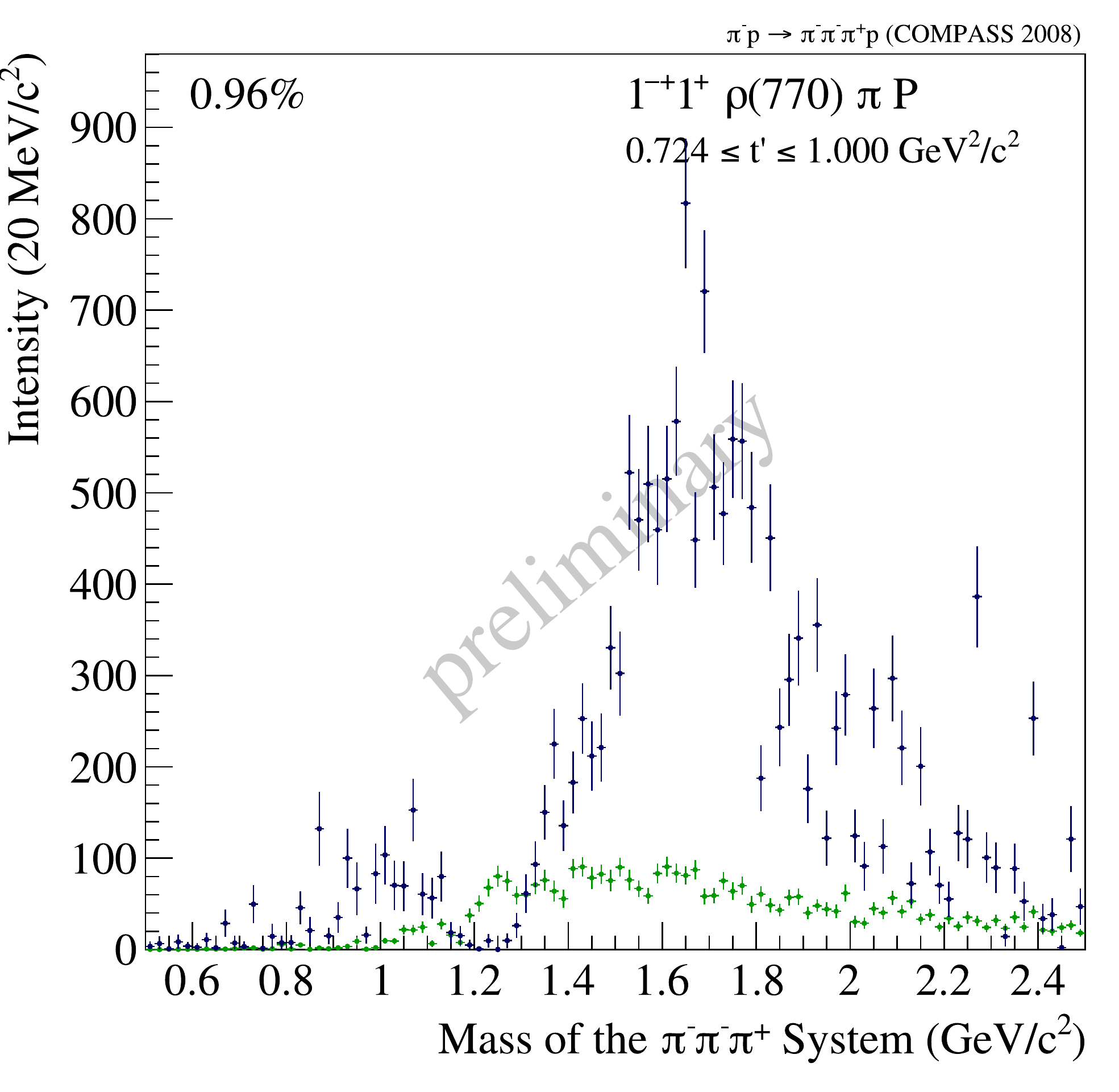}
  \caption{Intensity of the $1^{-+}\,1^+\,\rho(770)\,\pi\,P$ wave in
    different regions of $t'$ (left: low $t'$; right: high $t'$) for
    the $\pi^-\pi^+\pi^-$ final state (dark blue).  The partial-wave
    projections of Monte-Carlo data generated according to a model of
    the Deck effect are overlaid in green.}
  \label{fig:1mp_tbins}
\end{figure}

A remarkable change of the shape of the intensity spectrum of the
$1^{-+}\,1^+\,\rho(770)\,\pi\,P$ wave with $t'$ is observed (see
Fig.~\ref{fig:1mp_tbins}).  At values of $t'$ below about
0.3~$(\text{GeV}/c)^2$, we observe no indication of a resonance peak
around $m_{3\pi} = 1.6$~GeV/$c^2$, where we would expect the
$\pi_1(1600)$.  For the $t'$ bins in the interval from 0.449 to
$1.000~(\text{GeV}/c)^2$, the observed intensities exhibit a very
different shape as compared to the low-$t'$ region, with a peak
structure emerging at about 1.6~GeV/$c^2$ and the intensity at lower
masses disappearing rapidly with increasing $t'$.  This is in contrast
to clean resonance signals like the $a_2(1320)$ in the
$2^{++}\,1^+\,\rho(770)\,\pi\,D$ wave, which, as expected, do not
change their shape with $t'$.  The observed $t'$ behavior of the
$1^{-+}$ intensity is therefore a strong indication that non-resonant
contributions play a dominant role.  Interestingly, the relative
phases of the $1^{-+}\,1^+\,\rho(770)\,\pi\,P$ wave with respect to
other waves are virtually independent of $t'$.  As an example,
Fig.~\ref{fig:1mp} right shows the phase relative to the
$1^{++}\,0^+\,\rho(770)\,\pi\,S$ wave for four $t'$ bins.  In all
bins, a slowly rising phase with a total phase motion of about
60\textdegree\ is observed around the 1.6 ~GeV/$c^2$ region.

It is believed that the non-resonant contribution in the $1^{-+}$ wave
originates predominantly from the Deck effect, in which the incoming
beam pion dissociates into the isobar and an off-shell pion that
scatters off the target proton to become on-shell~\cite{deck}.  As a
first step towards a better understanding of the non-resonant
contribution, Monte-Carlo data were generated that are distributed
according to a model of the Deck effect.  The model employed here is
very similar to that used in Ref.~\cite{accmor_deck}.  The
partial-wave projection of these Monte Carlo data is shown as green
points in Fig.~\ref{fig:1mp_tbins}.  In order to compare the
intensities of real data and the Deck-model pseudo data, the Monte
Carlo data are scaled to the $t'$-summed intensity of the $1^{-+}$
wave as observed in real data.  At values of $t'$ below about
0.3~$(\text{GeV}/c)^2$, the intensity distributions of real data and
Deck Monte Carlo exhibit strong similarities suggesting that the
observed intensity might be saturated by the Deck effect.  Starting
from $t' \approx 0.4~(\text{GeV}/c)^2$, the spectral shapes for Deck
pseudo-data and real data deviate from each other with the differences
increasing towards larger values of $t'$.  This leaves room for a
potential resonance signal.  It should be noted, however, that the
Deck pseudo data contain no resonant contributions. Therefore,
potential interference effects between the resonant and non-resonant
amplitudes cannot be assessed in this simple approach.

\subsection{The $a_1(1420)$}

\begin{figure}[!tb]
  \centering
  \includegraphics[width=0.45\textwidth]{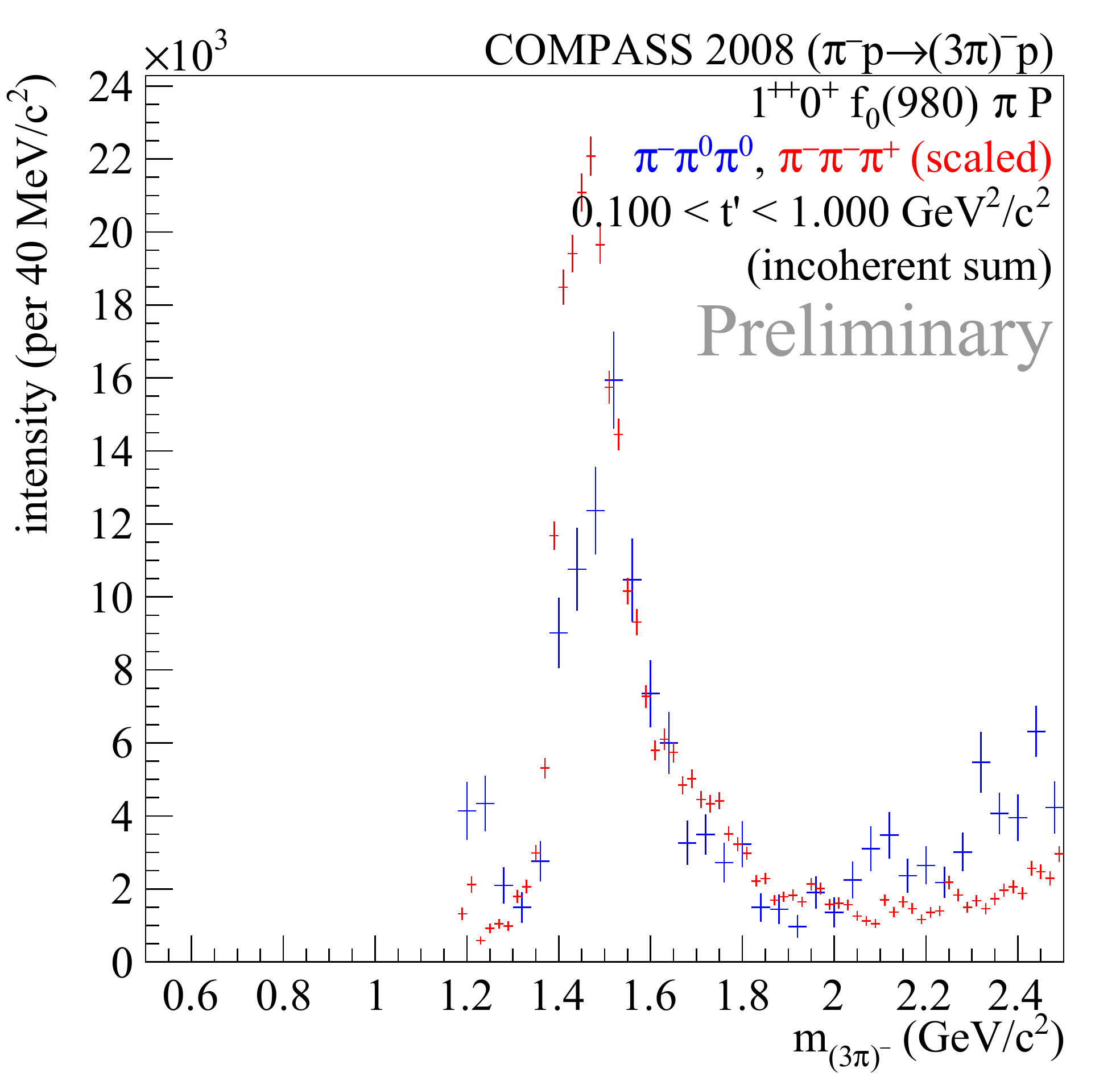}
  \includegraphics[width=0.45\textwidth]{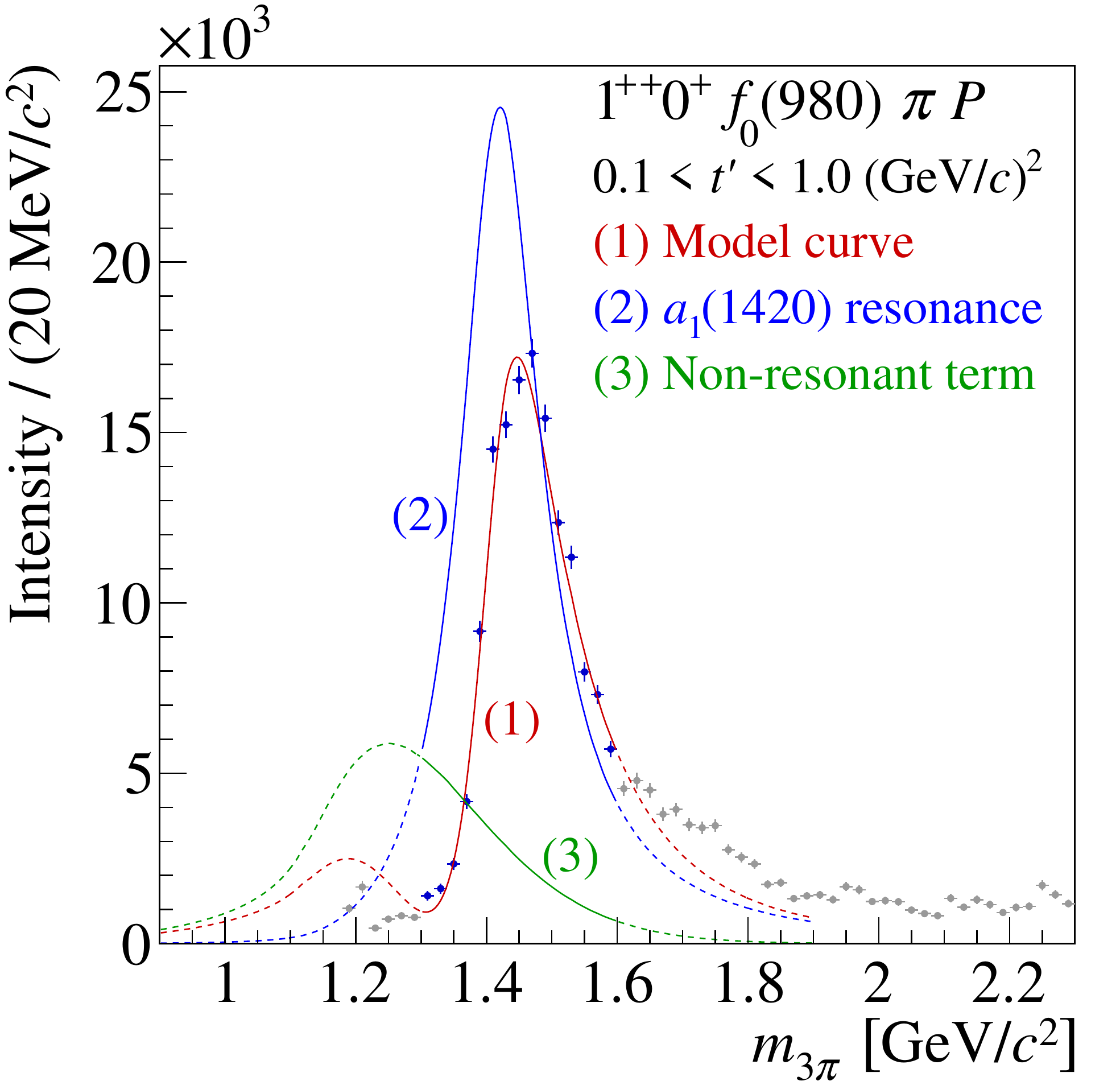}
  \caption{Left: Intensity of the $1^{++}\,0^+\,f_0(980)\,\pi\,P$ wave
    summed over all $t'$ bins for the $\pi^-\pi^+\pi^-$ (red) and the
    $\pi^-\pi^0\pi^0$ (blue) final states.  Right: Result of a
    resonance-model fit to the $\pi^-\pi^+\pi^-$ data~\cite{a1_1420}.
    The data points correspond to the red points in the left figure.}
  \label{fig:1pp_f0_intens}
\end{figure}

\begin{figure}[!tb]
  \centering
  \includegraphics[width=0.45\textwidth]{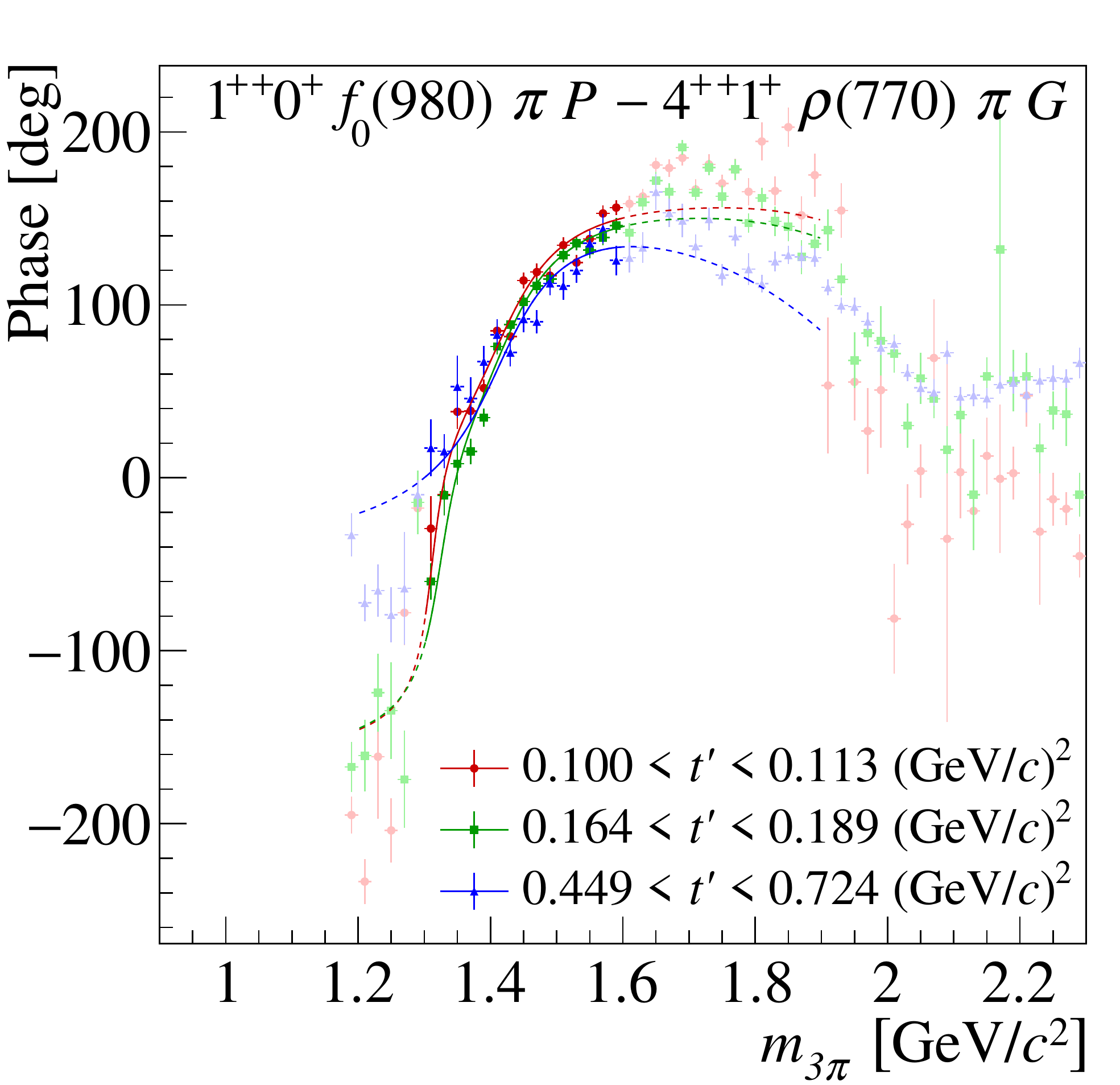}
  \includegraphics[width=0.45\textwidth]{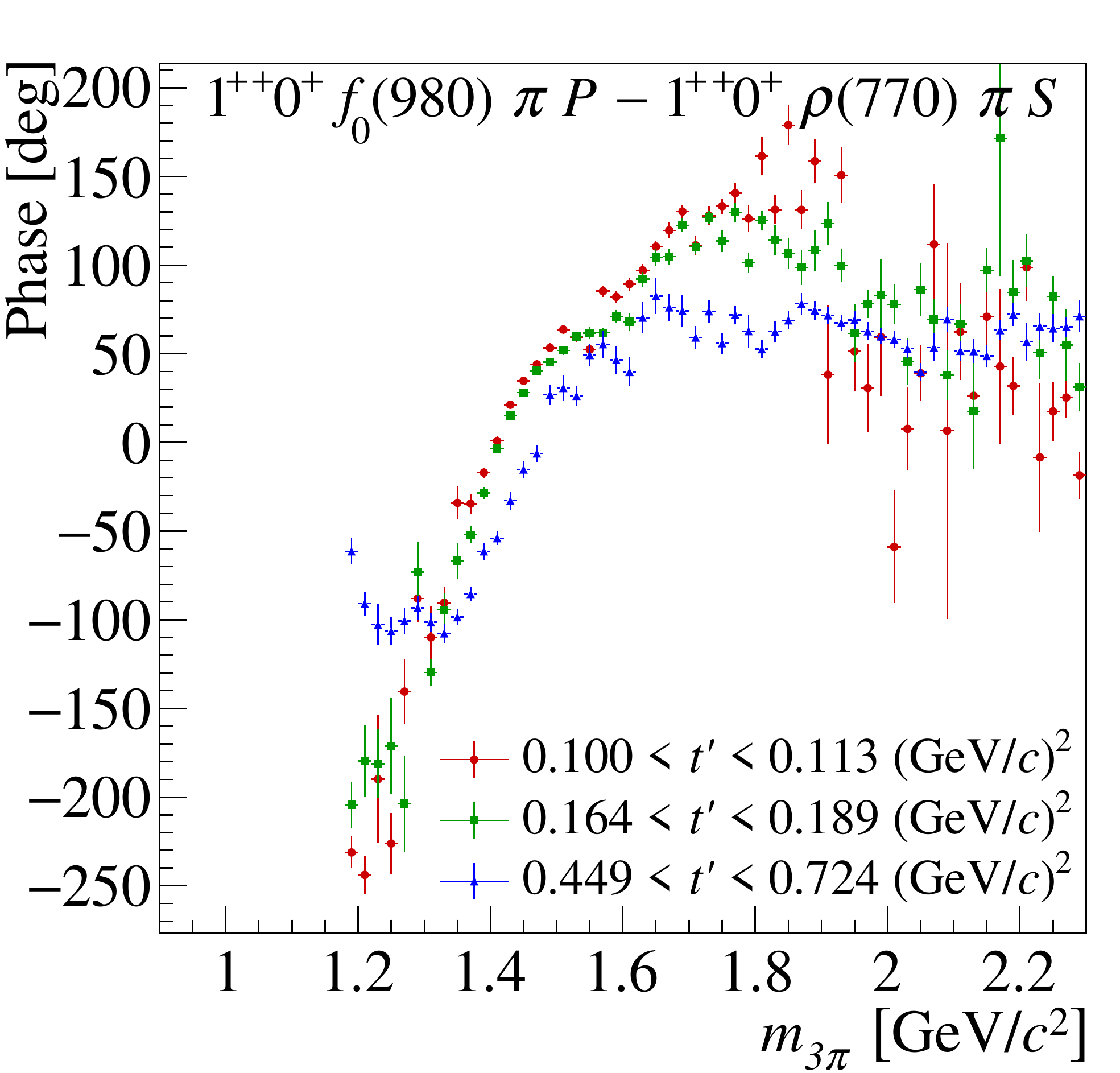}
  \caption{Examples for relative phases of the
    $1^{++}\,0^+\,f_0(980)\,\pi\,P$ wave with respect to the
    $4^{++}\,1^+\,\rho(770)\,\pi\,G$ (left) and the
    $1^{++}\,0^+\,\rho(770)\,\pi\,S$ wave (right).  The phases are
    shown for three different $t'$ regions indicated by the
    color. Figures from Ref.~\cite{a1_1420}.}
  \label{fig:1pp_f0_phase}
\end{figure}

A surprising find in the COMPASS data was a pronounced narrow peak at
about 1.4~GeV/$c^2$ in the $1^{++}\,0^+\,f_0(980)\,\pi\,P$ wave (see
Fig.~\ref{fig:1pp_f0_intens}).  The peak is observed with similar
shape in the $\pi^-\pi^+\pi^-$ and $\pi^-\pi^0\pi^0$ data and is
robust against variations of the PWA model.  In addition to the peak
in the partial-wave intensity, rapid phase variations with respect to
most waves are observed in the 1.4~GeV/$c^2$ region (see
Fig.~\ref{fig:1pp_f0_phase}).  The phase motion as well as the peak
shape change only little with $t'$.

In order to test the compatibility of the signal with a Breit-Wigner
resonance, a resonance-model fit was performed using a novel method,
where the intensities and relative phases of three waves
($1^{++}\,0^+\,f_0(980)\,\pi\,P$, $2^{++}\,1^+\,\rho(770)\,\pi\,D$,
and $4^{++}\,1^+\,\rho(770)\,\pi\,G$) were fit simultaneously in all
11 $t'$ bins~\cite{a1_1420}.  Forcing the resonance parameters to be
the same across all $t'$ bins leads to an improved separation of
resonant and non-resonant contribution as compared to previous
analyses that did not incorporate the $t'$ information.  The
Breit-Wigner model describes the peak in the
$1^{++}\,0^+\,f_0(980)\,\pi\,P$ wave well and yields a mass of
$m_0 = 1414^{+15}_{-13}$~MeV/$c^2$ and a width of
$\Gamma_0 = (153^{+8}_{-23})$~MeV/$c^2$ for the $a_1(1420)$.  Due to
the high statistical precision of the data, the uncertainties are
dominated by systematic effects.

The $a_1(1420)$ signal is remarkable in many ways.  It appears in a
mass region that is well studied since decades.  However, previous
experiments were unable to see the peak, because it contributes only
0.25\% to the total intensity.  The $a_1(1420)$ is very close in mass
to the $1^{++}$ ground state, the $a_1(1260)$.  But it has a much
smaller width than the $a_1(1260)$.  The $a_1(1420)$ peak is seen only
in the $f_0(980)\,\pi$ decay mode of the $1^{++}$ waves and lies
suspiciously close to the $K\, \bar{K}^*(892)$ threshold.

The nature of the $a_1(1420)$ is still unclear and several
interpretations were proposed.  It could be the isospin partner to the
$f_1(1420)$.  It was also described as a two-quark-tetraquark mixed
state~\cite{wang} and a tetraquark with mixed flavor
symmetry~\cite{chen}.  Other models do not require an additional
resonance: Ref.~\cite{berger1,berger2} proposes resonant re-scattering
corrections in the Deck process as an explanation, whereas
Ref.~\cite{bonn} suggests a branching point in the triangle
rescattering diagram for
$a_1(1260) \to K\, \bar{K}^*(892) \to K\, \bar{K}\, \pi \to f_0(980)\,
\pi$.
More detailed studies are needed in order to distinguish between these
models.

\subsection{Extraction of $\pi\pi$ $S$-Wave Isobar Amplitudes from Data}

The PWA of the $3\pi$ system is based on the isobar model, where fixed
amplitudes are used for the description of the $\pi^+\pi^-$
intermediate states.  However, we cannot exclude that the fit results
are biased by the employed isobar parametrizations.  This is true in
particular for the isoscalar $J^{PC} = 0^{++}$ isobars.  In the PWA
model, a broad $\pi\pi$ $S$-wave component is used, the
parametrization of which is extracted from $\pi\pi$ $S$-wave
elastic-scattering data~\cite{amp}.  In addition, the $f_0(980)$,
described by a Flatt\'e form~\cite{flatte}, and the $f_0(1500)$,
parametrized by a relativistic Breit-Wigner amplitude, are included as
isobars.  In order to study possible bias due to these
parametrizations and to ensure that the observed $a_1(1420)$ signal is
truly related to the narrow $f_0(980)$, a novel analysis
method~\cite{long_paper} inspired by Ref.~\cite{e791} was developed.
In this so-called \emph{freed-isobar} analysis, the three fixed
parametrizations for the $0^{++}$ isobar amplitudes are replaced by a
set of piecewise constant complex-valued functions that fully cover
the allowed two-pion mass range.  This way the whole $0^{++}$ isobar
amplitude is extracted as a function of the $3\pi$ mass.  In contrast
to the conventional isobar approach, which uses the same isobar
parametrization in different partial waves, the freed-isobar method
permits different isobar amplitudes for different intermediate states
$X^-$.  A more detailed description of the analysis method can be found
in Ref.~\cite{long_paper}.

\begin{figure}[!tb]
  \centering
  \begin{minipage}{\textwidth}
    \centering
    \includegraphics[width=0.45\textwidth]{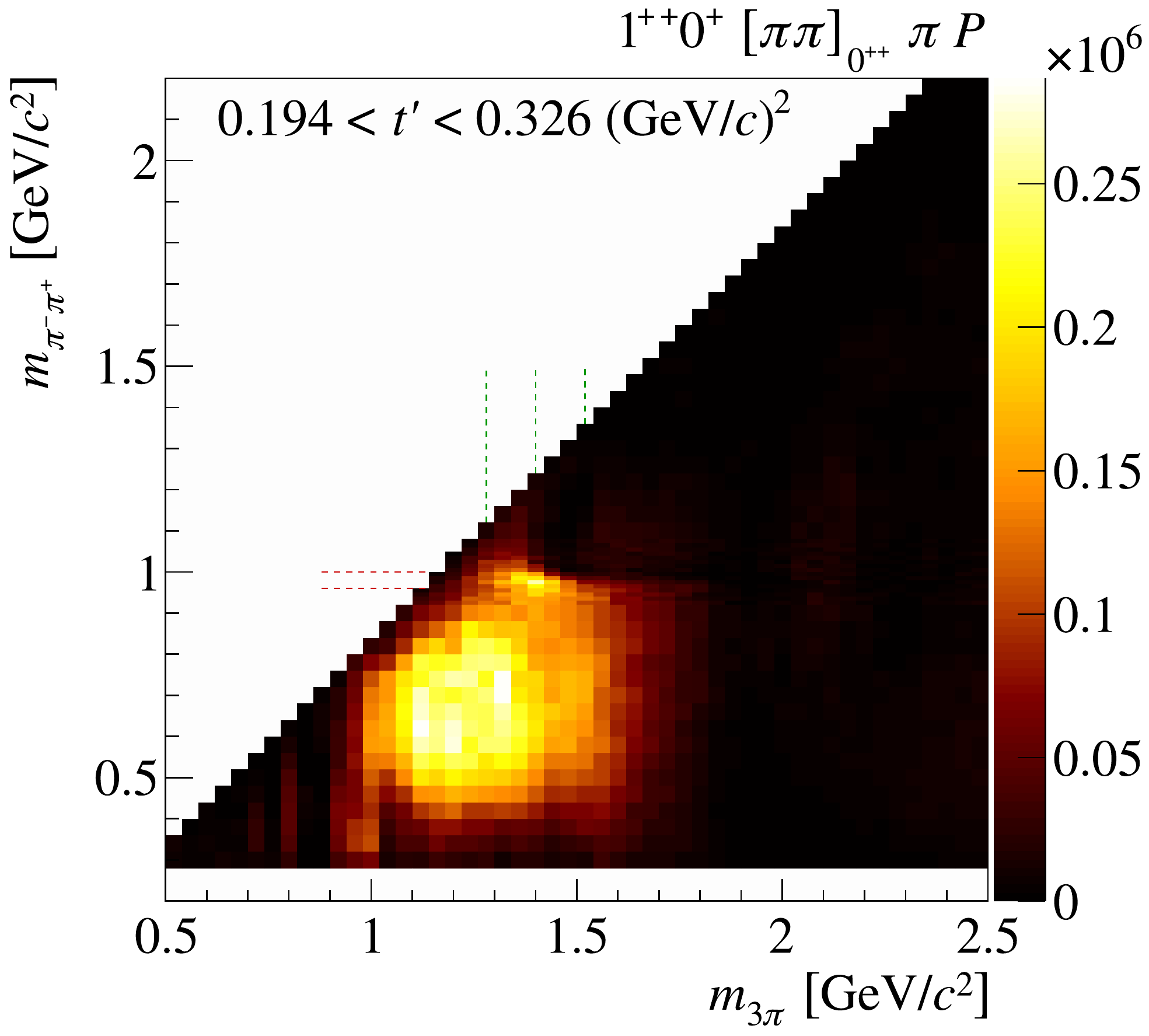} \\
    \includegraphics[width=0.45\textwidth]{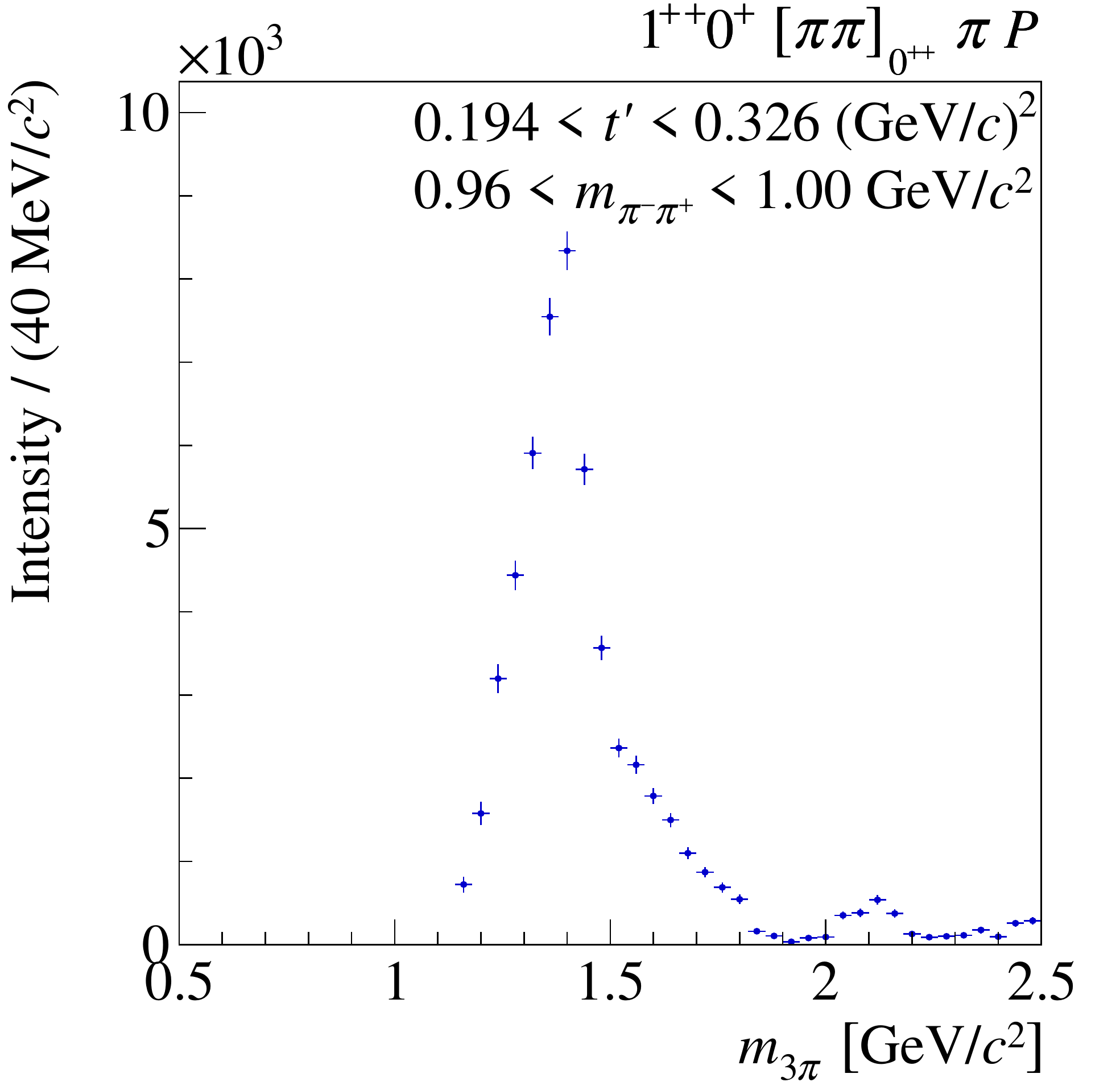}
    \includegraphics[width=0.45\textwidth]{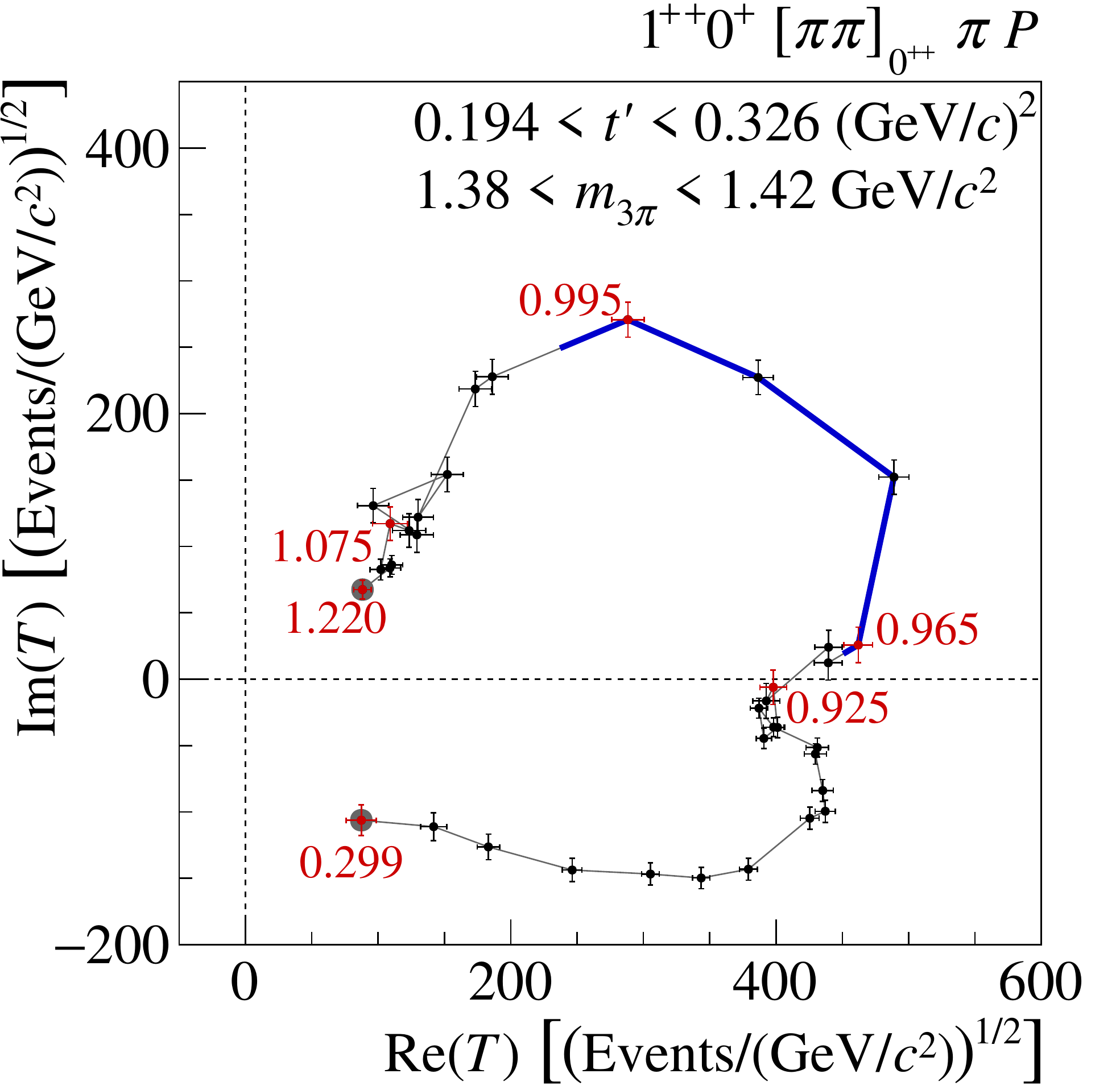}
  \end{minipage}
  \caption{Intensity of the $\pi\pi$ $S$-wave component of the
    $J^{PC}\,M^\epsilon = 1^{++}\,0^+$ partial wave resulting from the
    freed-isobar fit.  Top: two-dimensional representation of the
    partial-wave intensity as a function of $m_{\pi^+\pi^-}$ and
    $m_{3\pi}$.  Bottom left: intensity as a function of $m_{3\pi}$
    summed over the $m_{\pi^+\pi^-}$ interval around the $f_0(980)$
    indicated by the pair of horizontal dashed lines in the top
    figure.  Bottom right: Argand diagram representing the
    $m_{\pi^+\pi^-}$ dependence of the partial-wave amplitude for the
    $3\pi$ mass bin at the $a_1(1420)$ measured with respect to the
    $1^{++}\,0^+\,\rho(770)\,\pi\,S$ wave. Figures from
    Ref.~\cite{long_paper}.}
  \label{fig:freed-isobar}
\end{figure}

The freed-isobar method leads to a reduced model dependence and gives
additional information about the $\pi^+\pi^-$ subsystem at the cost of
a considerable increase in the number of free parameters in the PWA
fit.  Thus, even for large data sets, the freed-isobar approach can
only be applied to a subset of partial waves.  We performed a
freed-isobar PWA, where the fixed parametrizations of the broad
$\pi\pi$ $S$-wave component, of the $f_0(980)$, and of the $f_0(1500)$
were replaced by piece-wise constant isobar amplitudes for the $3\pi$
partial waves $0^{-+}\,0^+\,[\pi\pi]_{0^{++}}\,\pi\,S$,
$1^{++}\,0^+\,[\pi\pi]_{0^{++}}\,\pi\,P$, and
$2^{-+}\,0^+\,[\pi\pi]_{0^{++}}\,\pi\,D$.
Figure~\ref{fig:freed-isobar} top shows the two-dimensional intensity
distribution of the $1^{++}\,0^+\,[\pi\pi]_{0^{++}}\,\pi\,P$ wave as a
function of $m_{\pi^+\pi^-}$ and $m_{3\pi}$.  The distribution
exhibits a broad maximum around $m_{3\pi} = 1.2$~GeV/$c^2$ and between
0.6 and 0.8 ~GeV/$c^2$ in $m_{\pi^+\pi^-}$.  A smaller peak is
observed in the $f_0(980)$ region at $m_{3\pi} \approx 1.4$~GeV/$c^2$.
This peak is more obvious in Fig.~\ref{fig:freed-isobar} bottom left,
which shows the intensity distribution summed over the two-pion mass
interval around the $f_0(980)$ as indicated by the pair of horizontal
dashed lines in Fig.~\ref{fig:freed-isobar} top.  The peak is similar
in position and shape to the $a_1(1420)$ peak in the
$1^{++}\,0^+\,f_0(980)\,\pi\,P$ wave
(cf. Fig.~\ref{fig:1pp_f0_intens}).  The resonant nature of the
$f_0(980)$ becomes apparent in Fig.~\ref{fig:freed-isobar} bottom
right, which shows the $m_{\pi^+\pi^-}$ dependence of the extracted
amplitude at the $a_1(1420)$ peak in form of an Argand diagram.  The
phase is measured with respect to the $1^{++}\,0^+\,\rho(770)\,\pi\,S$
wave.  The $f_0(980)$ contribution shows up as a semicircle-like
structure (highlighted by the blue line) with a shifted origin.  This
demonstrates that the observed $a_1(1420)$ signal in the
$f_0(980)\,\pi$ decay mode is not an artifact of the $0^{++}$ isobar
parametrizations used in the conventional PWA method.  More results of
the freed-isobar PWA are discussed in Refs.~\cite{long_paper,fabian}.

\section{ACKNOWLEDGMENTS}

This work was supported by the BMBF, the MLL and the Cluster of
Excellence Exc153 ``Origin and Structure of the Universe''.

\bibliographystyle{aipnum-cp}%
\bibliography{bgrube}%

\end{document}